\documentclass[reprint,prl,showpacs,superscriptaddress,aps]{revtex4-1}
\usepackage{amsmath,amssymb,graphicx,microtype,color}
\usepackage{dcolumn}
\usepackage{bm}
\usepackage{hyperref}
\usepackage[mathlines]{lineno}
\usepackage{setspace}

\begin{document}

\preprint{APS/123-QED}

\title{Contractility-induced phase separation in active solids}

\author{Sifan Yin}
\affiliation{School of Engineering and Applied Sciences, Harvard University, Cambridge, MA 02138, USA}
\author{L.\ Mahadevan}
 \email{lmahadev@g.harvard.edu}
\affiliation{School of Engineering and Applied Sciences, Harvard University, Cambridge, MA 02138, USA}
\affiliation{Department of Organismic and Evolutionary Biology, Harvard University, Cambridge, MA 02138}
\affiliation{Department of Physics, Harvard University, Cambridge, MA 02138}

\begin{abstract}
A combination of cellular contractility and active phase separation in cell-matrix composites is thought to be an enabler of spatiotemporal patterning in multicellular tissues across scales, from somitogenesis to cartilage condensation. To characterize these phenomena,  we provide a general theory that incorporates active cellular contractility into the classical Cahn--Hilliard--Larch{\'e} model for phase separation in passive viscoelastic solids. We investigate the dynamics of phase separation in this model and show how a homogeneous mixture can be destabilized by activity via either a pitchfork or Hopf bifurcation, resulting in stable phase separation and/or traveling waves. Numerical simulations of the full equations allow us to track the evolution of the resulting self-organized patterns, in both periodic and mechanically constrained domains, and in different geometries.   Altogether, our study underscores the importance of integrating both cellular activity and mechanical phase separation in understanding patterning in soft, active biosolids, and might explain previous experimental observations of cartilage condensation in both in-vivo and in-vitro settings.

\end{abstract}
\pacs{}
\maketitle
Phase separation, long known to be an important driver in the patterning of material systems \cite{Onuki_book}, is increasingly being recognized to be an important pattern-forming mechanism that is also pervasive in biological systems across scales, from biomolecular condensates at the cellular scale \cite{Shin2017} to patterning of pollen grains \cite{Radja2021} to tissue morphogenesis in skin \cite{shyer2017emergent}, limbs \cite{Murray2003,Hiscock2017}, cartilage and bone \cite{Sala2011}. However, unlike in passive material systems where phase separation is driven by gradients in chemical potential, in living systems, molecular and cellular activity is an additional variable that can and does affect how patterns arise in space and time.

\begin{figure}[ht]
  \includegraphics[width=\linewidth]{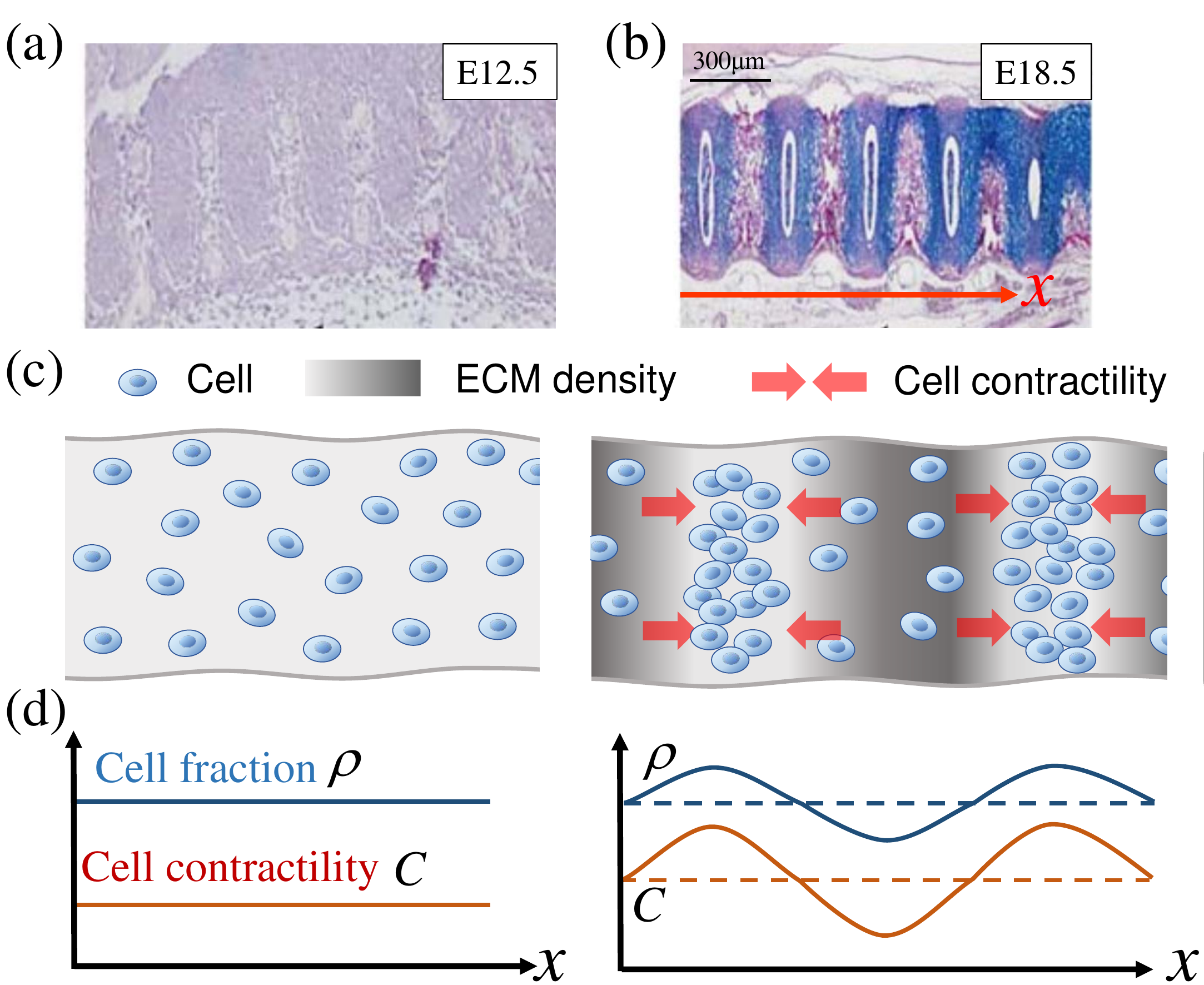}
   \vspace{-1.5em}
   \caption{(Color online) Illustration of biological phase separation process. (a-b) {\it In vivo} condensation of tracheal cartilage rings of mouse embryo at early(a) and late(b) stages \cite{Park2010}.(Scale bar: 300$\rm{\mu m}$ is estimated from \cite{Lin2013}) The blue and pink regions are mesenchymal cells and ECM, respectively. (c) Schematic of the cell-ECM mixture. Initially homogeneous soft cells (blue) scattered in ECM (gray) can contract and deform the viscoelastic biphasic mixture. When a local cell contractility rises above bifurcation threshold $C>C_{\rm cr}$, cells start to aggregate, which in turn leads to larger local contraction. Cellular contractility and movement are indicated by red arrows, and gray color map denotes local ECM fraction. (d) Profiles of the cell fraction $\rho$ and active contractility $C$ at homogeneous(left) and patterned(right) states.}
  \label{fig1} 
\end{figure}

In multicellular tissues, the focus of this letter, an experimentally well-studied example of  patterning that has many signatures of phase separation is that of pre-cartilage condensation \cite{Oster1983,Oster1985,Park2010,Sala2011}, an integrated multicellular process involving biochemical signal regulation, cell migration, cellular contractility, and extracellular matrix (ECM) deformation. 
Molecular manipulations show that eliminating active cellular contractility and/or motility prevents condensation {\it in-vivo} \cite{shyer2017emergent,Palmquist2022}. Furthermore reconstituted micromasses of cell and ECM show the formation of synthetic patterns that are accompanied by cell contractility and phase separation of the cell and ECM \cite{Lin2013,Saha2017}. Over the past few decades, various mechanisms have been proposed to explain patterning in these multi-phase systems, and fall into one of four broad categories:  Turing-like mechanisms based on differential diffusion, chemotaxis-based processes that focus on cell movement in response to chemokines, differential adhesion, and differential growth (see \cite{Iber2022} for a review). Here, guided by old and new experimental observations, we propose a new framework based on cellular contractility that can lead to phase separation and patterning. 

As a concrete example to set the stage, in Fig.~\ref{fig1}(a-b) we show the process of tracheal cartilage ring formation in a mouse embryo over a period of a few days \cite{Park2010}. An initially homogeneous mixture of (blue) cells and (pink) ECM from an early stage transitions over a period of 6 days and segregates into periodic patterns. 



In Fig.~\ref{fig1}(c), we show how condensation might be driven by cellular contractility in a biphasic mixture composed of a passive but stiff ECM (gray) within which are embedded soft active cells (blue). In their mesenchymal state, these cells can move, generate large traction forces and also secrete ECM. In a minimal picture \cite{Oster1983}, when individual cell contractility rises above a critical threshold (which we will calculate), the regions with slightly higher cell fractions get more compacted, leading to local cell aggregation which generates even larger contraction forces. This thus leads to a spatial separation of active cells and the passive ECM; however unlike in classical phase separation of and in liquid environments that continually coarsens till just a single phase boundary persists, here the balance between active cell contractility and the passive elastic response of the ECM eventually arrests coarsening, leading to non-trivial persistent patterns. We show this schematically in [Fig.~\ref{fig1}(c,d)] and now turn to quantify how this picture has the potential to explain the periodic patterns manifest in early cartilage patterns.



\paragraph{Active Mechanical Model}
\begin{figure*}[t]
   \includegraphics[width=\linewidth]{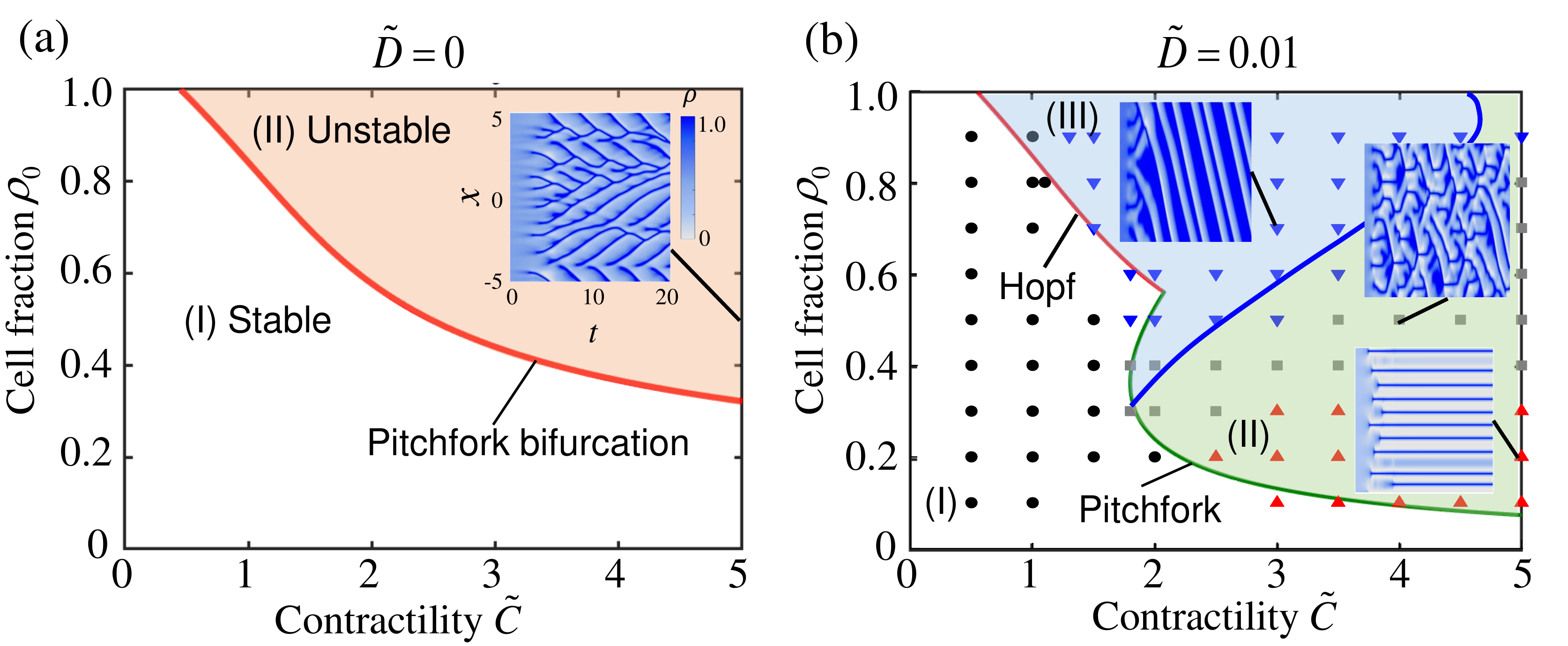}
   \vspace{-1em}
   \caption{\label{fig2} (Color online) Phase diagrams in the normalized $\rho_0-\tilde{C}$ plane obtained from linear stability analysis on 1D model, Eq.~\eqref{DispersionRelation}. Colored regions depict ${\rm Re}(\omega)>0$. (a) For diffusivity ${\tilde D}=0$, pink indicates ${\rm Im}(\omega)=0$ for all wave numbers. The inset shows a typical oscillatory pattern with nearly zero diffusivity ${\tilde D}=0.0001$ [Movie S1(a)] with color representing cell fraction $\rho$ from $0$ (gray) to $1$ (blue), and horizontal axis for time $t\in[0,20]\tau$ and vertical axis for $x\in[-5,5]\ell$. Other parameters are ${\tilde C}=5$ and $\rho_0=0.5$. (b) For non-zero diffusivity, $\tilde{D}=0.01$, three regions (I) stable (white), (II) stationary phase separation (green), and (III) oscillatory and traveling waves (blue) are distinguished from linear stability analysis, Eq. ~\eqref{DispersionRelation}. White region (I) represents all the modes being stable with $\text{Re}(\omega)<0$; Green region (II) represents a finite band of modes with $\text{Re}(\omega)>0$ whereas $\text{Im}(\omega)=0$; Blue region (III) represents a finite band of complex eigenvalues with positive real parts. Colored symbols show the systematic numerical results of the full 1D model, Eq.~\eqref{Non-equ}, with periodic boundary conditions. Black dots for no pattern formation; red up-triangles for phase separation [Movie S1(b)], and blue down-triangles for traveling waves [Movie S1(c)]. Additional complex and irregular patterns (gray squares) near the interface of regions (II) and (III) are observed, which exhibit two-direction traveling and continuous fission and merging [Movie S1(d)]. Three typical spatiotemporal plots of the cell fraction $\rho$ including traveling waves ($\tilde{C}=3$, $\rho_0=0.7$), stationary phase separation ($\tilde{C}=5$, $\rho_0=0.2$), and complex transition patterns ($\tilde{C}=4$, $\rho_0=0.5$) are shown in the insets with horizontal axis for time $t\in[0,100]\tau$ and vertical axis for $x\in[-5,5]\ell$.}
   \end{figure*}
Our theoretical framework for cartilage condensation follows from a simple generalization of the well-known formalism for passive phase separation \cite{Larche1973} to account for active processes. We start with a consideration of isotropic two-phase systems composed of an active phase representing mesenchymal cells that can generate active contraction, and a passive phase representing ECM, and model the mixture as a viscoelastic solid capable of phase separation and deformation. This bi-phasic system can be described in terms of two fields that vary in space-time: the cell fraction $\rho(x,t)$ and a displacement field $\bm{u}(x,t)$. Assuming no new production of the ECM, mass conservation  implies that the ECM fraction is $(1-\rho)$. In terms of the displacement field, we can also define the velocity field $\bm{v}(x,t) = \partial_t \bm{u}(x,t)$, the strain is $\bm \varepsilon= [\nabla {\bm u}+(\nabla {\bm u})^{\rm T}]/2$, and the strain rate is ${\dot{\bm \varepsilon}}= [\nabla{\bm v}+(\nabla {\bm v})^{\rm T}]/2$.  To model the combined effects of active cellular contraction, cell migration, mechanical deformation, and a thermodynamic driving force, we write the total free energy of the system as a generalization of the classical Cahn--Hilliard--Larch{\'e} model for phase separation \cite{Cahn1958,Cahn1961,Larche1973} that now reads as
\begin{equation}
  \mathcal{E}(\rho,{\bm u})=\int_{\Omega}\left[g_{\rm CH}(\rho)+ \frac12{\bm \sigma}^{\rm p}(\rho,{\bm \varepsilon},\dot{{\bm \varepsilon}}):{\bm \varepsilon} + {\bm \sigma}^{\rm a}(\rho):{\bm \varepsilon}\right]{\rm d} {\bm x},
\end{equation}
where $g_{\rm CH}(\rho)=\frac12\alpha\rho^2+\frac14\beta\rho^4+\frac12\gamma|\nabla \rho|^2 $ is the Ginzburg--Landau free energy with $\alpha<0$, $\beta>0$ defining the double-well potential, and $\gamma>0$ is the interfacial energy. The second and third terms in the expression above correspond to the mechanical strain energy due to the passive stress ${\bm \sigma}^{\rm p}$ and the work done by the active stress ${\bm \sigma}^{\rm a}$. The passive mechanical response of the ECM-cell composite is assumed to follow the Kelvin--Voigt model which for two/three-dimensional linear viscoelastic materials reads as \cite{Landau1970}
\begin{equation}\label{ConstitutiveEq}
   {\bm \sigma}^{\rm p}=\left[2G(\rho){\bm \varepsilon}+\lambda(\rho) \theta \bm{I}\right]+\left[2\mu_1(\rho)\dot{\bm \varepsilon}+\mu_2(\rho)\dot{\theta}\bm{I}\right],
\end{equation}
where  $\theta = \nabla\cdot {\bm u}$ is the dilation, and $\dot{\theta} = \nabla\cdot {\bm v}$ is the dilation rate. Here, the mechanical parameters of the mixture are the shear modulus $G$, first Lam{\'e} coefficient $\lambda$, shear viscosity $\mu_1$ and bulk viscosity $\mu_2$.  Following linear mixture theory \cite{Christensen_book}, the mechanical properties are taken as a weighted average of each phase, i.e., $\phi(\rho)=\phi_{\rm a}\rho+\phi_{\rm p}(1-\rho)$, where $\phi=E,~ \zeta,~G,~\lambda,~\mu_1,~\mu_2$ and the subscript `a' and `p' represent the active and passive phase, respectively.   \par

The active stress ${\bm \sigma}^{\rm a}$, which can depend on many biophysical variables \cite{Weber2018,Alonso2016,Radszuweit2013}, is assumed to be contractile and isotropic, and of the form ${\bm \sigma}^{\rm a}(\rho)=Cf(\rho)\bm{I}$, where $C>0$ is the contractile amplitude, and the function $f(\rho)$ has a saturating Hill-type functional form \cite{Murray2003, Mietke2019}
\begin{equation}
 Cf(\rho)=C\frac{\rho^2}{1+K\rho^2},
\end{equation}
where the contractility monotonically increases with cell fraction $\rho\in [0,1]$ within a regime $[0,C/(1+K)]$.\par

Then, given the form of the active stress and the constitutive form for the solid,  we may write the equations for mass and momentum balance equations (neglecting the effects of inertia), accounting for material convection and thermodynamic driving force, as
\begin{equation}\label{MassConservation}
  \partial_t \rho = -\nabla \cdot(\rho \bm{v})+D\nabla^2\left[\frac{\delta \mathcal{E}}{\delta \rho}\right],
\end{equation}
\begin{equation}\label{ForceBalance}
-\frac{\delta \mathcal{E}}{\delta \bm{u}}=\nabla \cdot ({\bm\sigma}^{\rm p}+Cf(\rho)\bm{I})=\eta \bm{v}.
\end{equation}
Here $D$ is the cellular diffusivity,  $\delta \mathcal{E}/\delta \rho=\mu$ is the chemo-mechanical potential, and $\eta$ is the effective frictional coefficient associated with movement relative to a background. The current framework complements previous work \cite{Weber2018,Mietke2019,Bois2011,Radszuweit2013} which focused on differential activity in such examples as a biphasic continuum, active fluid surfaces etc., while here we account for contractility-driven phase separation in a viscoelastic continuum model of multicellular tissue-ECM mixtures.\par

\paragraph{Linear stability analysis}

To illustrate the essential aspects of the active biphasic system, we start with a simple one-dimensional problem, (\cite{SM} section I. Eq.~S4), where the mechanical parameters are the elastic modulus $E$ and viscosity $\zeta$. After rescaling the governing variables using a characteristic length ${\ell}=\sqrt{\zeta_{\rm p}/\eta}$ and time $\tau=\zeta_{\rm p} / E_{\rm p}$, we can write the mass and force balance equations Eq.\eqref{MassConservation} and Eq.\eqref{ForceBalance} as
\begin{subequations}\label{Non-equ}
\begin{align}
  \partial_t \rho&=-\partial_x(\rho v)+\tilde{D}\partial_{xx}\left[\tilde{\alpha}\rho+\tilde{\beta}\rho^3-\tilde{\gamma}\rho_{xx}\right.\notag\\
  &~~~\left.+\tilde{C}f'(\rho)u_x+\frac12(\tilde{E}_{\rm a}-1)u_x^2+\frac12(\tilde{\zeta}_{\rm a}-1)u_x v_x \right],\label{Non-equ1}
  \\
 v&=\partial_x\left[\tilde{E}(\rho)u_{x}+\tilde{\zeta}(\rho)v_x+\tilde{C}f(\rho)\right],\label{Non-equ2}
  \end{align}
 \end{subequations}
where the dimensionless physical quantities are denoted with an upper tilde (see \cite{SM} Section I. Table S2). The dimensionless parameters in the model include the scaled domain size $\tilde{L}=L/\ell$, the ratio of elastic moduli $\tilde{E}_{\rm a}=E_{\rm a}/E_{\rm p}$, the ratio of viscosities $\tilde{\zeta}_{\rm a}=\zeta_{\rm a}/\zeta_{\rm p}$, the chemical potential parameters $\tilde{\alpha}$, $\tilde{\beta}$, and $\tilde{\gamma}$, as well as the scaled diffusivity $\tilde{D}=D\eta$, and the scaled contractility $\tilde{C}=C/E_{\rm p}$. Of these, the critical parameters are the last two (see \cite{SM} Section II.A for details of the other parameters).\par

We next carry out a linear stability analysis of the homogeneous steady state solutions of (6) with a uniform fraction of active cells $\rho=\rho_0$ in a stationary medium $u=v=0$. Assuming small perturbations of the active cell fraction and displacement written as a superposition of Fourier modes $(\delta \rho,\delta u)\propto\ {\rm exp}\left[ \omega_n t+ik_nx\right]$, where $\omega_n$ is the growth rate and $k_n=2\pi n/\tilde{L}$ is the wave number of each mode $n$, we substitute these  into Eqs.\eqref{Non-equ1}-\eqref{Non-equ2}. This yields the dispersion relation $\omega_n=\omega(k_n^2)$ as 
\begin{equation}\label{DispersionRelation}
  \omega(k_n^2)= \frac{1}{2a(k_n^2)}\left[-b(k_n^2)\pm \sqrt{b(k_n^2)^2-4a(k_n^2)c(k_n^2)}\right],
\end{equation}
with $b(k_n^2)$ and $c(k_n^2)$ are algebraically complex functions of the wavenumber [\cite{SM} section II. Eq.~S17], and $a(k_n^2)=[1+{\tilde \zeta}(\rho_o)k_n^2]/k_n^2>0$. We see that instability sets in when  $b(k_n^2)<0$ or $c(k_n^2)<0$.\par

In the absence of diffusivity, i.e. when $\tilde{D}=0$, increasing contractility can trigger a supercritical pitchfork bifurcation. To understand this (see \cite{SM} Section II. Eq.~S15) we see that the displacement follows the simple equation $\partial_t \delta u= [{\tilde E}(\rho_0)-\rho_0{\tilde C}f'(\rho_0)]/a(k_n^2) \delta u$. Thus small perturbations will grow/decay depending on the balance between the stabilizing elastic and destabilizing active stresses; when the scaled activity $\tilde{C}>\tilde{C}_{\rm cr}=\tilde{E}(\rho_0)/[\rho_0 f'(\rho_0)]$, the growth rate is purely real and positive for all wave numbers. We note that ${\tilde C}_{\rm cr}$ decreases monotonically with the initial (uniform) cell fraction $\rho_0$ [Fig.~\ref{fig2}(a), red line]. The dominant wavenumber at the onset of the instability is given by (see \cite{SM} section II, Eqs. S17--S21) 
\begin{equation}\label{ScalingLaw}
    k_{\rm cr}^2=(1/{\tilde\gamma})\left[{\tilde C}^2f'(\rho_0)^2/{\tilde E}(\rho_0)-({\tilde \alpha}+3{\tilde \beta}\rho_0^2)\right].
\end{equation}
We see that active contraction favors small wavelengths while the elasticity prefers long wavelengths, but in either case the critical wavenumber is inversely proportional to the interfacial energy, i.e. $k_{\rm cr}\sim 1/\sqrt{\tilde{\gamma}}$ (see \cite{SM} Section IV Fig. S5).\par

With non-vanishing diffusivity of the active cells, either a Hopf and pitchfork bifurcation can occur. In Fig.~\ref{fig2}(b), we see three quantitatively different regions in the phase diagram spanned by $\rho_0$ and $\tilde C$: region I is pattern-free, region II corresponds to unstable modes with purely real eigenvalues (stationary phase separation), and region III corresponds to unstable modes with complex eigenvalues (traveling waves of phase separation). Furthermore, we see that the critical contractility $\tilde{C}_{\rm cr}$ for a pitchfork bifurcation when $\tilde{D}=0$ [see Fig.~\ref{fig2}(a) and (b)], is larger than that when $\tilde{D} \ne 0$ because the driving force due to the Cahn--Hilliard chemical potential promotes phase separation and lowers the required active contractility. However, the critical wavenumber for both cases is same as Eq.~\eqref{ScalingLaw} (see \cite{SM} section II.A Eqs. S23--S29).

To understand the role of material rheology on the nature of the instability, we now turn to a comparison of the purely elastic, purely viscous, and viscoelastic mixture systems. The purely viscous mixture with active contractility is always unstable via a pitchfork bifurcation, because the linear perturbation amplitude satisfies the equation $a(k_n^2)\delta u_{tt}+b(k_n^2)\delta u_t+c(k_n^2)\delta u=0$ with $a(k_n^2)>0$ and $c(k_n^2)<0$ which has at least one positive real growth rate (see \cite{SM}, section II.C). For the purely elastic and the viscoelastic mixture, while elasticity helps to stabilize the system, contractility serves to destabilize it via either a pitchfork or a Hopf bifurcation [see \cite{SM} Fig. S2(a)]. Finally, we note that in the absence of contractility, the system only exhibits a pitchfork bifurcation associated with the classical picture of phase separation in the Cahn--Hilliard system.

\paragraph{Numerical simulation} To follow the dynamics of pattern evolution beyond the onset of the instability, we numerically solved the nonlinear equations (6a,b) after introducing a numerical inertia term $\Gamma\ddot{u}$ in a one-dimensional domain using a pseudo-spectral method \cite{Trefethen2000} (spatial and temporal convergence are validated by increasing the number of collocation points and time steps [see \cite{SM} section III.C Fig. S4]) using biologically plausible parameters (see \cite{SM} Table S1 and S2). 
A phase diagram for the different patterns as a function of the initial active cell fraction $\rho_0$ and the scaled contractility $\tilde{C}$ obtained from systematic numerical simulations shown in Fig.~\ref{fig2}(b) agrees well with our linear instability analysis,  and goes further in describing complex dynamic behaviors  away from the onset of instability.\par

When the diffusivity $\tilde{D} = 0$ [Fig.~\ref{fig2}(a) inset], in the unstable region (II) we see the continuous formation and merger of bi-directional short waves of active cellular condensation [Movie S1(a)], with fission and fusion statistically balancing each other. 
{Increasing diffusivity} yields three regimes as shown in Fig.~\ref{fig2}(b); no pattern (black dots), stationary phase separation [red up triangles, Movie S1(b)], and  travelling waves [blue down triangles, Movie S1(c)]. Near the boundary of the transitions, the dynamics of patterns can be irregular and chaotic, as for example when wave fronts merge into one and then move collectively [Fig.~\ref{fig2}(b), middle inset, Movie S1(d)]. Comparing  the patterns obtained from our numerical simulations with those of linear stability analysis, we find that the wavelength $\lambda=\ell$ is close to but a little smaller than the linear prediction $\lambda=2\pi\ell/k\approx 1.4\ell$ [see \cite{SM} Fig. S1(c)]. Furthermore, the numerical characteristic width of periodic stripe pattern (wavelength) shows $\lambda\sim \sqrt{\tilde \gamma}$ independent of diffusivity [see \cite{SM} Fig. S5], consistent with the linear stability analysis result given by Eq.~\eqref{ScalingLaw}.\par

\begin{figure}[t]
 \includegraphics[width=\linewidth]{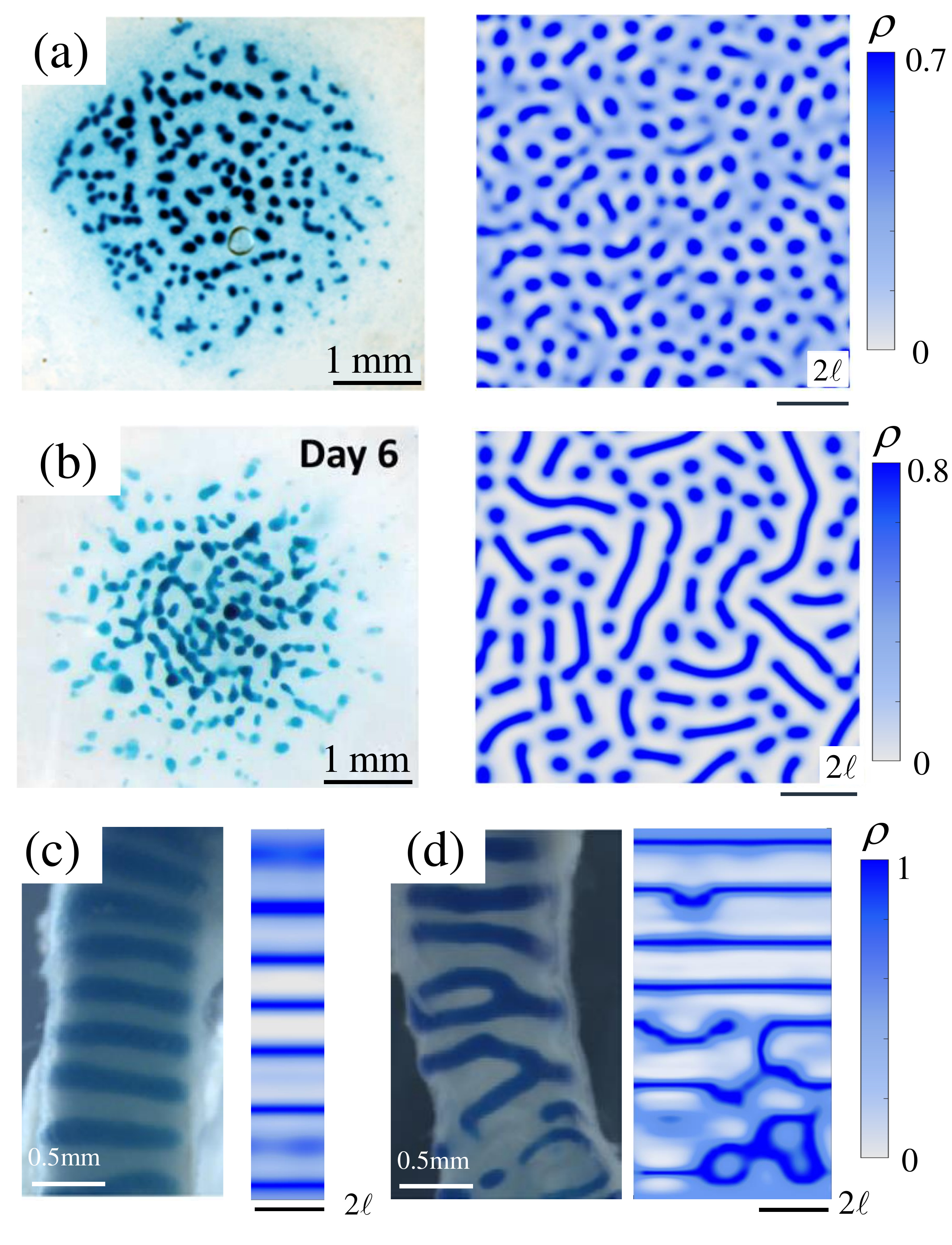}
  \vspace{-1.5em}
\caption{\label{fig3} Experimental and numerical simulations of representative two-dimensional dynamic patterns.(a-b) Left column shows the images of typical {\it in vitro} spots(a) \cite{Saha2017} and stripes(b) \cite{Butterfield2017} patterns of pre-cartilage condensations in micromasses cultures of mesenchymal cells from mouse embryos. Right column exhibits 2D simulations of spots and labyrinth.  (c--d) Experimental normal(c) and abnormal(d) tracheal cartilage rings \cite{Sala2011} and numerical simulations. The parameters used are (a) $\tilde{C}=10$, $\tilde{D}=0.001$, $\rho_0=0.3$, $\tilde{\gamma}=0.01$ at $\tilde{t}=50$, periodic BCs; (b) $\tilde{C}=10$, $\tilde{D}=0.01$, $\rho_0=0.3$, $\tilde{\gamma}=0.01$ at $\tilde{t}=25$, periodic BCs. (c-d) Constant velocity $\tilde{v}=0.001\ell/\tau$ on top boundary with $\tilde{C}=5$,$\tilde{D}=0.01$, $\rho_0=0.5$, $\tilde{\gamma}=0.1$, ${\tilde L}_y=10\ell$, ${\tilde L}_x=2\ell$(c) and $5\ell$(d). Snapshots are captured at $\tilde{t}=100$.}
\end{figure}

Two-dimensional simulations exhibit more complex but qualitatively similar behaviors to those seen in our one-dimensional simulations. Using periodic boundary conditions and random initial conditions, we simulated the dynamics of Eq. (4,5) (in the parameter space for phase separation, corresponding to region II in Fig.~\ref{fig2}b). Our results show how ordered and disordered spots and stripes (labyrinthine patterns) arise, similar to the experimentally observed pre-cartilage condensation patterns in micromasses cultures shown in Fig.~\ref{fig3}(a) and (b) \cite{Saha2017,Butterfield2017}, and Movie S2. Using the parameters $\tilde \alpha= -1$, $\tilde \beta=1$, and $\tilde \gamma=0.01-0.2$, and carrying out a Fourier analysis of our simulations yields the dominant (dimensionless) wavenumbers for spots as $k_{\rm spot}=8.19$(a) and stripes as $k_{\rm stripe}=7.11$(b). Using experimental estimates \cite{Saha2017,Butterfield2017} for $\ell\approx 0.5{\rm mm}$, we observe reasonable agreement between experimental observations and numerical results for the wavelength of spots and stripes (see \cite{SM} section IV, Fig.~S6). \par

Finally, to investigate the role of boundary conditions to understand aspects of periodic cartilage condensation along the tubular trachea in mice \cite{Sala2011}, reproduced in Fig.~\ref{fig3}(c,d), we prescribe either a constant velocity or a constant displacement applied at the tube ends and periodic boundary conditions on the lateral boundaries associated with a cylindrical tube. Here, we neglect the effects of tube curvature by assuming that it is small relative to the wavelength of the patterns (and check that this is indeed the case later). In Fig.~\ref{fig3}(c), we see that our simulations  reproduce the experimentally observed periodic condensation patterns for low loading rates (or prescribed boundary displacements). Increasing the lateral dimensions or the rate of loading causes the pattern to become disordered as shown in Fig. 3d {(see \cite{SM} section V. Fig.~S7 and S8)}. 

\paragraph{Conclusions}
By accounting for the role of active contractility in a multiphase biological mixture of cells and ECM, we have shown how patterns can arise spontaneously in a homogeneous soft active solid, providing a framework to explain aspects of tissue morphogenesis that are consistent with old and new observations of condensation in multicellular tissues such as cartilage.  
The relatively simple generalization of classical Cahn--Hilliard--Larch{\'e} theory that accounts for forces that arise from activity as well as those from passive mechanical deformation and thermodynamics shows how different stationary and dynamic patterns including spots and stripes, traveling and oscillatory waves and complex transition patterns arise naturally, in one and two-dimensional periodic geometries. By accounting for different mechanical boundary conditions, we also see the appearance of both ordered and disordered patterns. 

Natural next steps include linking our theory to experiments that probe the role of activity, cell and ECM fraction and geometry of the tissue, and generalizing the theory to account for tissue growth and activity profiles that vary in space-time and are coupled to gene expression levels. 


\providecommand{\noopsort}[1]{}\providecommand{\singleletter}[1]{#1}%
%



\begin{thebibliography}{27}%
\makeatletter
\providecommand \@ifxundefined [1]{%
 \@ifx{#1\undefined}
}%
\providecommand \@ifnum [1]{%
 \ifnum #1\expandafter \@firstoftwo
 \else \expandafter \@secondoftwo
 \fi
}%
\providecommand \@ifx [1]{%
 \ifx #1\expandafter \@firstoftwo
 \else \expandafter \@secondoftwo
 \fi
}%
\providecommand \natexlab [1]{#1}%
\providecommand \enquote  [1]{``#1''}%
\providecommand \bibnamefont  [1]{#1}%
\providecommand \bibfnamefont [1]{#1}%
\providecommand \citenamefont [1]{#1}%
\providecommand \href@noop [0]{\@secondoftwo}%
\providecommand \href [0]{\begingroup \@sanitize@url \@href}%
\providecommand \@href[1]{\@@startlink{#1}\@@href}%
\providecommand \@@href[1]{\endgroup#1\@@endlink}%
\providecommand \@sanitize@url [0]{\catcode `\\12\catcode `\$12\catcode
  `\&12\catcode `\#12\catcode `\^12\catcode `\_12\catcode `\%12\relax}%
\providecommand \@@startlink[1]{}%
\providecommand \@@endlink[0]{}%
\providecommand \url  [0]{\begingroup\@sanitize@url \@url }%
\providecommand \@url [1]{\endgroup\@href {#1}{\urlprefix }}%
\providecommand \urlprefix  [0]{URL }%
\providecommand \Eprint [0]{\href }%
\providecommand \doibase [0]{https://doi.org/}%
\providecommand \selectlanguage [0]{\@gobble}%
\providecommand \bibinfo  [0]{\@secondoftwo}%
\providecommand \bibfield  [0]{\@secondoftwo}%
\providecommand \translation [1]{[#1]}%
\providecommand \BibitemOpen [0]{}%
\providecommand \bibitemStop [0]{}%
\providecommand \bibitemNoStop [0]{.\EOS\space}%
\providecommand \EOS [0]{\spacefactor3000\relax}%
\providecommand \BibitemShut  [1]{\csname bibitem#1\endcsname}%
\let\auto@bib@innerbib\@empty
\bibitem [{\citenamefont {Onuki}(2002)}]{Onuki_book}%
  \BibitemOpen
  \bibfield  {author} {\bibinfo {author} {\bibfnamefont {A.}~\bibnamefont
  {Onuki}},\ }\href {https://doi.org/10.1017/CBO9780511534874} {\emph {\bibinfo
  {title} {Phase Transition Dynamics}}}\ (\bibinfo  {publisher} {Cambridge
  University Press},\ \bibinfo {year} {2002})\BibitemShut {NoStop}%
\bibitem [{\citenamefont {Shin}\ and\ \citenamefont
  {Brangwynne}(2017)}]{Shin2017}%
  \BibitemOpen
  \bibfield  {author} {\bibinfo {author} {\bibfnamefont {Y.}~\bibnamefont
  {Shin}}\ and\ \bibinfo {author} {\bibfnamefont {C.~P.}\ \bibnamefont
  {Brangwynne}},\ }\href {https://doi.org/10.1126/science.aaf4382} {\bibfield
  {journal} {\bibinfo  {journal} {Science}\ }\textbf {\bibinfo {volume}
  {357}},\ \bibinfo {pages} {eaaf4382} (\bibinfo {year} {2017})}\BibitemShut
  {NoStop}%
\bibitem [{\citenamefont {Radja}(2021)}]{Radja2021}%
  \BibitemOpen
  \bibfield  {author} {\bibinfo {author} {\bibfnamefont {A.}~\bibnamefont
  {Radja}},\ }\href@noop {} {\bibfield  {journal} {\bibinfo  {journal} {J. Exp.
  Zool. B Mol. Dev. Evol.}\ }\textbf {\bibinfo {volume} {336}},\ \bibinfo
  {pages} {629} (\bibinfo {year} {2021})}\BibitemShut {NoStop}%
\bibitem [{\citenamefont {Shyer}\ \emph {et~al.}(2017)\citenamefont {Shyer},
  \citenamefont {Rodrigues}, \citenamefont {Schroeder}, \citenamefont
  {Kassianidou}, \citenamefont {Kumar},\ and\ \citenamefont
  {Harland}}]{shyer2017emergent}%
  \BibitemOpen
  \bibfield  {author} {\bibinfo {author} {\bibfnamefont {A.~E.}\ \bibnamefont
  {Shyer}}, \bibinfo {author} {\bibfnamefont {A.~R.}\ \bibnamefont
  {Rodrigues}}, \bibinfo {author} {\bibfnamefont {G.~G.}\ \bibnamefont
  {Schroeder}}, \bibinfo {author} {\bibfnamefont {E.}~\bibnamefont
  {Kassianidou}}, \bibinfo {author} {\bibfnamefont {S.}~\bibnamefont {Kumar}},\
  and\ \bibinfo {author} {\bibfnamefont {R.~M.}\ \bibnamefont {Harland}},\
  }\href@noop {} {\bibfield  {journal} {\bibinfo  {journal} {Science}\ }\textbf
  {\bibinfo {volume} {357}},\ \bibinfo {pages} {811} (\bibinfo {year}
  {2017})}\BibitemShut {NoStop}%
\bibitem [{\citenamefont {Murray}(2003)}]{Murray2003}%
  \BibitemOpen
  \bibfield  {author} {\bibinfo {author} {\bibfnamefont {J.}~\bibnamefont
  {Murray}},\ }\href {https://doi.org/10.1007/b98869} {\emph {\bibinfo {title}
  {Mathematical Biology II: Spatial Models and Biomedical Applications}}},\
  \bibinfo {series} {Interdisciplinary Applied Mathematics}, Vol.~\bibinfo
  {volume} {18}\ (\bibinfo  {publisher} {Springer New York},\ \bibinfo {year}
  {2003})\BibitemShut {NoStop}%
\bibitem [{\citenamefont {Hiscock}\ \emph {et~al.}(2017)\citenamefont
  {Hiscock}, \citenamefont {Tschopp},\ and\ \citenamefont
  {Tabin}}]{Hiscock2017}%
  \BibitemOpen
  \bibfield  {author} {\bibinfo {author} {\bibfnamefont {T.~W.}\ \bibnamefont
  {Hiscock}}, \bibinfo {author} {\bibfnamefont {P.}~\bibnamefont {Tschopp}},\
  and\ \bibinfo {author} {\bibfnamefont {C.~J.}\ \bibnamefont {Tabin}},\ }\href
  {https://doi.org/10.1016/j.devcel.2017.04.021} {\bibfield  {journal}
  {\bibinfo  {journal} {Dev. Cell}\ }\textbf {\bibinfo {volume} {41}},\
  \bibinfo {pages} {459} (\bibinfo {year} {2017})}\BibitemShut {NoStop}%
\bibitem [{\citenamefont {Sala}\ \emph {et~al.}(2011)\citenamefont {Sala},
  \citenamefont {Del~Moral}, \citenamefont {Tiozzo}, \citenamefont {Alam},
  \citenamefont {Warburton}, \citenamefont {Grikscheit}, \citenamefont
  {Veltmaat},\ and\ \citenamefont {Bellusci}}]{Sala2011}%
  \BibitemOpen
  \bibfield  {author} {\bibinfo {author} {\bibfnamefont {F.~G.}\ \bibnamefont
  {Sala}}, \bibinfo {author} {\bibfnamefont {P.~M.}\ \bibnamefont {Del~Moral}},
  \bibinfo {author} {\bibfnamefont {C.}~\bibnamefont {Tiozzo}}, \bibinfo
  {author} {\bibfnamefont {D.~A.}\ \bibnamefont {Alam}}, \bibinfo {author}
  {\bibfnamefont {D.}~\bibnamefont {Warburton}}, \bibinfo {author}
  {\bibfnamefont {T.}~\bibnamefont {Grikscheit}}, \bibinfo {author}
  {\bibfnamefont {J.~M.}\ \bibnamefont {Veltmaat}},\ and\ \bibinfo {author}
  {\bibfnamefont {S.}~\bibnamefont {Bellusci}},\ }\href
  {https://doi.org/10.1242/dev.051680} {\bibfield  {journal} {\bibinfo
  {journal} {Development}\ }\textbf {\bibinfo {volume} {138}},\ \bibinfo
  {pages} {273} (\bibinfo {year} {2011})}\BibitemShut {NoStop}%
\bibitem [{\citenamefont {Park}\ \emph {et~al.}(2010)\citenamefont {Park},
  \citenamefont {Zhang}, \citenamefont {Choi}, \citenamefont {Trinh},\ and\
  \citenamefont {Kim}}]{Park2010}%
  \BibitemOpen
  \bibfield  {author} {\bibinfo {author} {\bibfnamefont {J.}~\bibnamefont
  {Park}}, \bibinfo {author} {\bibfnamefont {J.~J.}\ \bibnamefont {Zhang}},
  \bibinfo {author} {\bibfnamefont {R.}~\bibnamefont {Choi}}, \bibinfo {author}
  {\bibfnamefont {I.}~\bibnamefont {Trinh}},\ and\ \bibinfo {author}
  {\bibfnamefont {P.~C.}\ \bibnamefont {Kim}},\ }\href
  {https://doi.org/10.1007/s11626-009-9255-9} {\bibfield  {journal} {\bibinfo
  {journal} {In Vitro Cell. Dev. Biol. Anim.}\ }\textbf {\bibinfo {volume}
  {46}},\ \bibinfo {pages} {92} (\bibinfo {year} {2010})}\BibitemShut {NoStop}%
\bibitem [{\citenamefont {Lin}\ \emph {et~al.}(2014)\citenamefont {Lin},
  \citenamefont {Tzen}, \citenamefont {Lee}, \citenamefont {Smith},
  \citenamefont {Campbell},\ and\ \citenamefont {Chen}}]{Lin2013}%
  \BibitemOpen
  \bibfield  {author} {\bibinfo {author} {\bibfnamefont {S.-S.}\ \bibnamefont
  {Lin}}, \bibinfo {author} {\bibfnamefont {B.-H.}\ \bibnamefont {Tzen}},
  \bibinfo {author} {\bibfnamefont {K.-R.}\ \bibnamefont {Lee}}, \bibinfo
  {author} {\bibfnamefont {R.~J.~H.}\ \bibnamefont {Smith}}, \bibinfo {author}
  {\bibfnamefont {K.~P.}\ \bibnamefont {Campbell}},\ and\ \bibinfo {author}
  {\bibfnamefont {C.-C.}\ \bibnamefont {Chen}},\ }\href
  {https://doi.org/10.1073/pnas.1323112111} {\bibfield  {journal} {\bibinfo
  {journal} {Proc. Natl. Acad. Sci. U.S.A.}\ }\textbf {\bibinfo {volume}
  {111}},\ \bibinfo {pages} {E1990} (\bibinfo {year} {2014})}\BibitemShut
  {NoStop}%
\bibitem [{\citenamefont {Oster}\ \emph {et~al.}(1983)\citenamefont {Oster},
  \citenamefont {Murray},\ and\ \citenamefont {Harris}}]{Oster1983}%
  \BibitemOpen
  \bibfield  {author} {\bibinfo {author} {\bibfnamefont {G.~F.}\ \bibnamefont
  {Oster}}, \bibinfo {author} {\bibfnamefont {J.~D.}\ \bibnamefont {Murray}},\
  and\ \bibinfo {author} {\bibfnamefont {A.~K.}\ \bibnamefont {Harris}},\
  }\href@noop {} {\bibfield  {journal} {\bibinfo  {journal} {J. Embryol. Exp.
  Morphol.}\ }\textbf {\bibinfo {volume} {78}},\ \bibinfo {pages} {83}
  (\bibinfo {year} {1983})}\BibitemShut {NoStop}%
\bibitem [{\citenamefont {Oster}\ \emph {et~al.}(1985)\citenamefont {Oster},
  \citenamefont {Murray},\ and\ \citenamefont {Maini}}]{Oster1985}%
  \BibitemOpen
  \bibfield  {author} {\bibinfo {author} {\bibfnamefont {G.~F.}\ \bibnamefont
  {Oster}}, \bibinfo {author} {\bibfnamefont {J.~D.}\ \bibnamefont {Murray}},\
  and\ \bibinfo {author} {\bibfnamefont {P.~K.}\ \bibnamefont {Maini}},\
  }\href@noop {} {\bibfield  {journal} {\bibinfo  {journal} {J. Embryol. Exp.
  Morphol.}\ }\textbf {\bibinfo {volume} {89}},\ \bibinfo {pages} {93}
  (\bibinfo {year} {1985})}\BibitemShut {NoStop}%
\bibitem [{\citenamefont {Palmquist}\ \emph {et~al.}(2022)\citenamefont
  {Palmquist}, \citenamefont {Tiemann}, \citenamefont {Ezzeddine},
  \citenamefont {Yang}, \citenamefont {Pfeifer}, \citenamefont {Erzberger},
  \citenamefont {Rodrigues},\ and\ \citenamefont {Shyer}}]{Palmquist2022}%
  \BibitemOpen
  \bibfield  {author} {\bibinfo {author} {\bibfnamefont {K.~H.}\ \bibnamefont
  {Palmquist}}, \bibinfo {author} {\bibfnamefont {S.~F.}\ \bibnamefont
  {Tiemann}}, \bibinfo {author} {\bibfnamefont {F.~L.}\ \bibnamefont
  {Ezzeddine}}, \bibinfo {author} {\bibfnamefont {S.}~\bibnamefont {Yang}},
  \bibinfo {author} {\bibfnamefont {C.~R.}\ \bibnamefont {Pfeifer}}, \bibinfo
  {author} {\bibfnamefont {A.}~\bibnamefont {Erzberger}}, \bibinfo {author}
  {\bibfnamefont {A.~R.}\ \bibnamefont {Rodrigues}},\ and\ \bibinfo {author}
  {\bibfnamefont {A.~E.}\ \bibnamefont {Shyer}},\ }\href@noop {} {\bibfield
  {journal} {\bibinfo  {journal} {Cell}\ } (\bibinfo {year}
  {2022})}\BibitemShut {NoStop}%
\bibitem [{\citenamefont {Saha}\ \emph {et~al.}(2017)\citenamefont {Saha},
  \citenamefont {Rolfe}, \citenamefont {Carroll}, \citenamefont {Kelly},\ and\
  \citenamefont {Murphy}}]{Saha2017}%
  \BibitemOpen
  \bibfield  {author} {\bibinfo {author} {\bibfnamefont {A.}~\bibnamefont
  {Saha}}, \bibinfo {author} {\bibfnamefont {R.}~\bibnamefont {Rolfe}},
  \bibinfo {author} {\bibfnamefont {S.}~\bibnamefont {Carroll}}, \bibinfo
  {author} {\bibfnamefont {D.~J.}\ \bibnamefont {Kelly}},\ and\ \bibinfo
  {author} {\bibfnamefont {P.}~\bibnamefont {Murphy}},\ }\href
  {https://doi.org/10.1007/s00441-016-2512-9} {\bibfield  {journal} {\bibinfo
  {journal} {Cell Tissue Res.}\ }\textbf {\bibinfo {volume} {368}},\ \bibinfo
  {pages} {47} (\bibinfo {year} {2017})}\BibitemShut {NoStop}%
\bibitem [{\citenamefont {Iber}\ and\ \citenamefont {M.}(2022)}]{Iber2022}%
  \BibitemOpen
  \bibfield  {author} {\bibinfo {author} {\bibfnamefont {D.}~\bibnamefont
  {Iber}}\ and\ \bibinfo {author} {\bibfnamefont {M.}~\bibnamefont {M.}},\
  }\href@noop {} {\bibfield  {journal} {\bibinfo  {journal} {Front. Cell Dev.
  Biol.}\ ,\ \bibinfo {pages} {900447}} (\bibinfo {year} {2022})}\BibitemShut
  {NoStop}%
\bibitem [{\citenamefont {Larch{\'e'}}\ and\ \citenamefont
  {Cahn}(1973)}]{Larche1973}%
  \BibitemOpen
  \bibfield  {author} {\bibinfo {author} {\bibfnamefont {F.}~\bibnamefont
  {Larch{\'e'}}}\ and\ \bibinfo {author} {\bibfnamefont {J.}~\bibnamefont
  {Cahn}},\ }\href
  {https://doi.org/https://doi.org/10.1016/0001-6160(73)90021-7} {\bibfield
  {journal} {\bibinfo  {journal} {Acta Metall.}\ }\textbf {\bibinfo {volume}
  {21}},\ \bibinfo {pages} {1051} (\bibinfo {year} {1973})}\BibitemShut
  {NoStop}%
\bibitem [{\citenamefont {Cahn}\ and\ \citenamefont
  {Hilliard}(1958)}]{Cahn1958}%
  \BibitemOpen
  \bibfield  {author} {\bibinfo {author} {\bibfnamefont {J.~W.}\ \bibnamefont
  {Cahn}}\ and\ \bibinfo {author} {\bibfnamefont {J.~E.}\ \bibnamefont
  {Hilliard}},\ }\href@noop {} {\bibfield  {journal} {\bibinfo  {journal} {J.
  Chem. Phys.}\ }\textbf {\bibinfo {volume} {28}},\ \bibinfo {pages} {258}
  (\bibinfo {year} {1958})}\BibitemShut {NoStop}%
\bibitem [{\citenamefont {Cahn}(1961)}]{Cahn1961}%
  \BibitemOpen
  \bibfield  {author} {\bibinfo {author} {\bibfnamefont {J.~W.}\ \bibnamefont
  {Cahn}},\ }\href@noop {} {\bibfield  {journal} {\bibinfo  {journal} {Acta
  Metall.}\ }\textbf {\bibinfo {volume} {9}},\ \bibinfo {pages} {795} (\bibinfo
  {year} {1961})}\BibitemShut {NoStop}%
\bibitem [{\citenamefont {Landau}\ \emph {et~al.}(1970)\citenamefont {Landau},
  \citenamefont {Lif{\v{s}}ic}, \citenamefont {Sykes},\ and\ \citenamefont
  {Reid}}]{Landau1970}%
  \BibitemOpen
  \bibfield  {author} {\bibinfo {author} {\bibfnamefont {E.}~\bibnamefont
  {Landau}, \bibfnamefont {L.D.}}, \bibinfo {author} {\bibfnamefont
  {E.}~\bibnamefont {Lif{\v{s}}ic}}, \bibinfo {author} {\bibfnamefont
  {J.}~\bibnamefont {Sykes}},\ and\ \bibinfo {author} {\bibfnamefont
  {W.}~\bibnamefont {Reid}},\ }\href@noop {} {\emph {\bibinfo {title} {Theory
  of Elasticity}}},\ \bibinfo {series} {Course of theoretical physics}\ No.\
  \bibinfo {number} {v. 7}\ (\bibinfo  {publisher} {Pergamon Press},\ \bibinfo
  {year} {1970})\BibitemShut {NoStop}%
\bibitem [{\citenamefont {Christensen}(2012)}]{Christensen_book}%
  \BibitemOpen
  \bibfield  {author} {\bibinfo {author} {\bibfnamefont {R.~M.}\ \bibnamefont
  {Christensen}},\ }\href@noop {} {\emph {\bibinfo {title} {Mechanics of
  composite materials}}}\ (\bibinfo  {publisher} {Courier Corporation},\
  \bibinfo {year} {2012})\BibitemShut {NoStop}%
\bibitem [{\citenamefont {Weber}\ \emph {et~al.}(2018)\citenamefont {Weber},
  \citenamefont {Rycroft},\ and\ \citenamefont {Mahadevan}}]{Weber2018}%
  \BibitemOpen
  \bibfield  {author} {\bibinfo {author} {\bibfnamefont {C.~A.}\ \bibnamefont
  {Weber}}, \bibinfo {author} {\bibfnamefont {C.~H.}\ \bibnamefont {Rycroft}},\
  and\ \bibinfo {author} {\bibfnamefont {L.}~\bibnamefont {Mahadevan}},\ }\href
  {https://doi.org/10.1103/PhysRevLett.120.248003} {\bibfield  {journal}
  {\bibinfo  {journal} {Phys. Rev. Lett.}\ }\textbf {\bibinfo {volume} {120}},\
  \bibinfo {pages} {248003} (\bibinfo {year} {2018})}\BibitemShut {NoStop}%
\bibitem [{\citenamefont {Alonso}\ \emph {et~al.}(2016)\citenamefont {Alonso},
  \citenamefont {Strachauer}, \citenamefont {Radszuweit}, \citenamefont
  {Bär},\ and\ \citenamefont {Hauser}}]{Alonso2016}%
  \BibitemOpen
  \bibfield  {author} {\bibinfo {author} {\bibfnamefont {S.}~\bibnamefont
  {Alonso}}, \bibinfo {author} {\bibfnamefont {U.}~\bibnamefont {Strachauer}},
  \bibinfo {author} {\bibfnamefont {M.}~\bibnamefont {Radszuweit}}, \bibinfo
  {author} {\bibfnamefont {M.}~\bibnamefont {Bär}},\ and\ \bibinfo {author}
  {\bibfnamefont {M.~J.~B.}\ \bibnamefont {Hauser}},\ }\href
  {https://doi.org/10.1016/j.physd.2015.09.017} {\bibfield  {journal} {\bibinfo
   {journal} {Physica D}\ }\textbf {\bibinfo {volume} {318--319}},\ \bibinfo
  {pages} {58} (\bibinfo {year} {2016})}\BibitemShut {NoStop}%
\bibitem [{\citenamefont {Radszuweit}\ \emph {et~al.}(2013)\citenamefont
  {Radszuweit}, \citenamefont {Alonso}, \citenamefont {Engel},\ and\
  \citenamefont {Bar}}]{Radszuweit2013}%
  \BibitemOpen
  \bibfield  {author} {\bibinfo {author} {\bibfnamefont {M.}~\bibnamefont
  {Radszuweit}}, \bibinfo {author} {\bibfnamefont {S.}~\bibnamefont {Alonso}},
  \bibinfo {author} {\bibfnamefont {H.}~\bibnamefont {Engel}},\ and\ \bibinfo
  {author} {\bibfnamefont {M.}~\bibnamefont {Bar}},\ }\href
  {https://doi.org/10.1103/PhysRevLett.110.138102} {\bibfield  {journal}
  {\bibinfo  {journal} {Phys. Rev. Lett.}\ }\textbf {\bibinfo {volume} {110}},\
  \bibinfo {pages} {138102} (\bibinfo {year} {2013})}\BibitemShut {NoStop}%
\bibitem [{\citenamefont {Mietke}\ \emph {et~al.}(2019)\citenamefont {Mietke},
  \citenamefont {Julicher},\ and\ \citenamefont {Sbalzarini}}]{Mietke2019}%
  \BibitemOpen
  \bibfield  {author} {\bibinfo {author} {\bibfnamefont {A.}~\bibnamefont
  {Mietke}}, \bibinfo {author} {\bibfnamefont {F.}~\bibnamefont {Julicher}},\
  and\ \bibinfo {author} {\bibfnamefont {I.~F.}\ \bibnamefont {Sbalzarini}},\
  }\href {https://doi.org/10.1073/pnas.1810896115} {\bibfield  {journal}
  {\bibinfo  {journal} {Proc. Natl. Acad. Sci. U.S.A.}\ }\textbf {\bibinfo
  {volume} {116}},\ \bibinfo {pages} {29} (\bibinfo {year} {2019})}\BibitemShut
  {NoStop}%
\bibitem [{\citenamefont {Bois}\ \emph {et~al.}(2011)\citenamefont {Bois},
  \citenamefont {Julicher},\ and\ \citenamefont {Grill}}]{Bois2011}%
  \BibitemOpen
  \bibfield  {author} {\bibinfo {author} {\bibfnamefont {J.~S.}\ \bibnamefont
  {Bois}}, \bibinfo {author} {\bibfnamefont {F.}~\bibnamefont {Julicher}},\
  and\ \bibinfo {author} {\bibfnamefont {S.~W.}\ \bibnamefont {Grill}},\ }\href
  {https://doi.org/10.1103/PhysRevLett.106.028103} {\bibfield  {journal}
  {\bibinfo  {journal} {Phys. Rev. Lett.}\ }\textbf {\bibinfo {volume} {106}},\
  \bibinfo {pages} {028103} (\bibinfo {year} {2011})}\BibitemShut {NoStop}%
\bibitem [{SM()}]{SM}%
  \BibitemOpen
  \href@noop {} {}\bibinfo {note} {See Supplemental Material for videos and
  more information}\BibitemShut {NoStop}%
\bibitem [{\citenamefont {Trefethen}(2000)}]{Trefethen2000}%
  \BibitemOpen
  \bibfield  {author} {\bibinfo {author} {\bibfnamefont {L.~N.}\ \bibnamefont
  {Trefethen}},\ }\href@noop {} {\emph {\bibinfo {title} {Spectral methods in
  MATLAB}}}\ (\bibinfo  {publisher} {SIAM},\ \bibinfo {year}
  {2000})\BibitemShut {NoStop}%
\bibitem [{\citenamefont {Butterfield}\ \emph {et~al.}(2017)\citenamefont
  {Butterfield}, \citenamefont {Qian},\ and\ \citenamefont
  {Logan}}]{Butterfield2017}%
  \BibitemOpen
  \bibfield  {author} {\bibinfo {author} {\bibfnamefont {N.~C.}\ \bibnamefont
  {Butterfield}}, \bibinfo {author} {\bibfnamefont {C.}~\bibnamefont {Qian}},\
  and\ \bibinfo {author} {\bibfnamefont {M.~P.~O.}\ \bibnamefont {Logan}},\
  }\href {https://doi.org/10.1371/journal.pone.0180453} {\bibfield  {journal}
  {\bibinfo  {journal} {PLoS One}\ }\textbf {\bibinfo {volume} {12}},\ \bibinfo
  {pages} {e0180453} (\bibinfo {year} {2017})}\BibitemShut {NoStop}%
\end{thebibliography}
\end{document}